# Discovery of the compact X-ray source inside the Cygnus Loop

Emi Miyata,[*] Hiroshi Tsunemi[*] Ken'ichi Torii,[†] and Kiyoshi Hashimotodani

*Department of Earth and Space Science, Graduate School of Science, Osaka University*

*1-1, Machikaneyama, Toyonaka, Osaka 560-0043*

*E-mail(EM) : miyata @ ess.sci.osaka-u.ac.jp*

Takeshi Tsuru[*] and Katsuji Koyama[*]

*Cosmic Ray Group, Department of Physics, Kyoto University*

*Kitashirakawa-Oiwake-Cho, Sakyo, Kyoto 606-8502*

Kazuya Ayani

*Bisei Astronomical Observatory*

*Ohkura, Bisei, Okayama 714-1411*

Kouji Ohta

*Department of Astronomy, Faculty of Science, Kyoto University*

*Sakyo-ku, Kyoto 606-8502*

and

Michitoshi Yoshida

*Okayama Astrophysical Observatory*

*Kamogata-cho, Asakuchi-gun, Okayama 719-02*

**Abstract**

We detected an X-ray compact source inside the Cygnus Loop during the observation project of the whole Cygnus Loop with the ASCA GIS. The source intensity is 0.11 c s$^{-1}$ for GIS and 0.15 c s$^{-1}$ for SIS, which is the strongest in the ASCA band. The X-ray spectra are well fitted by a power law spectrum of a photon index of $-2.1 \pm 0.1$ with neutral H column of

$(3.1 \pm 0.6) \times 10^{21} \mathrm{cm}^{-2}$. Taking into account the interstellar absorption feature, this source is X-ray bright mainly above 1 keV suggesting either an AGN or a rotating neutron star. So far, we did not detect intensity variation nor coherent pulsation mainly due to the limited observation time. There are several optical bright stellar objects within the error region of the X-ray image. We carried out the optical spectroscopy for the brightest source (V=+12.6) and found it to be a G star. The follow up deep observation both in optical and in X-ray wavelengths are strongly required.

1. Introduction

The supernova (SN) explosion is a source of heavy elements in the galaxy. After the explosion, they are left as supernova remnants (SNRs) which are bright in various wavelengths. The X-ray spectrum of the SNR is useful to perform the diagnostic of the plasma which contains heavy elements. The young SNRs, like Cas-A (Holt et al. 1996), Tycho (Hwang & Gotthelf 1997), and Kepler (Decourchelle et al. 1997), show various emission lines in their spectra. They are mainly originated from the ejecta rather than the interstellar matter (ISM). Whereas, no associated compact source is reported for those young SNRs showing thin thermal emission. There are three historical SNe containing an associated compact source inside: Crab nebula (SN1054), G11.2-0.3 (SN386), and 3C58 (SN1181) (Clark & Stephenson 1977). The young SNRs associated with compact sources usually show non-thermal emission.

Some of the middle aged SNRs containing compact sources show thermal emission: the Vela SNR (Kahn et al. 1983; Bocchino, Maggio, & Sciortino 1994), and Puppis-A (Petre, Becker, & Winkler 1996). The Cygnus Loop is one of the well studied middle aged SNRs in various wavelengths. There is a high temperature component in its center showing center filled structure above 2 keV (Hatsukade & Tsunemi 1990). Miyata et al. (1998) detected strong emission lines of Si, S, and Fe from the central part of the Cygnus Loop showing high metal abundance. Whereas, Miyata et al. (1994) reported the sub-solar abundance from the North-East limb of the Loop which must correspond to the abundance of the ISM. Taking into account the projection effect, Miyata et al. (1998) concluded that there were Si, S, and Fe rich plasmas in the center of the Loop. The metal rich plasma must have originated from the ejecta of the SN with the progenitor mass of 25 $M_\odot$, strongly suggesting the presence of a stellar remnant. Thorsett et al. (1994) reported that they discovered a pulsar (PSR J2043+2740) at 430 MHz which was 21 pc away from the center. This location is out of the shell. The distance inferred from the dispersion measure is consistent with that of the Cygnus Loop, whereas the characteristic age is $1.21 \times 10^6$ yr (Ray et al. 1996), which is two orders of magnitude longer than that of the Cygnus Loop ($1.8 \times 10^4$ yr; Ku et al. 1984). The transverse velocity (3,000 km s$^{-1}$) is also too high compared with other pulsars (Lyne & Lorimer 1994). Therefore, they ruled out the association between PSR J2043+2740 and the Cygnus Loop. So far, no compact source associated with the Cygnus Loop is established.

∗ CREST, Japan Science and Technology Corporation (JST)

† Present address: NASDA TKSC SURP, 2-1-1 Sengen, Tsukuba, Ibaraki, 305-8505 Japan

We have performed the observation project of the whole Cygnus Loop by using the ASCA GIS (Tanaka, Inoue, & Holt 1994). We present here the discovery of an X-ray compact source which is inside the Cygnus Loop.

## 2. Observation and data analysis

We started the observation project to cover the whole Cygnus Loop with the ASCA GIS. It was started in the PV phase observation in 1993 and was completed in the AO5 observation in 1997. So far we performed 28 pointing observations and each observation time was about 10 ks. In the observation performed on June 3, 1997, we found a compact source which was hardly seen in the ROSAT all-sky survey (Aschenbach 1994).

### 2.1. Observation

The source was detected both in the GISs (GIS-2 & GIS-3) and in the SIS (SIS-1). The SISs were operated in 1–CCD Faint mode, and each sensor observed different sky area. The source was detected at the edge of the SIS-1 FOV whereas it was not detected in the SIS-0. The GISs were operated in PH mode with nominal bit assignment (10–8–8–5–0–0). Figure 1(a) shows the X-ray surface brightness map obtained with the ROSAT all-sky survey (Aschenbach 1994) and the black circle shows the GIS FOV. The X-ray image obtained with the GIS is shown in figure 1(b). We found two compact sources within the GIS FOV. Source-1 shown in figure 1(b) has already detected both with the Einstein Observatory (Ku et al. 1984) and with the ROSAT as shown in figure 1(a). It has been identified as a K5 star in the Hipparcos catalogue. On the other hand, the brightest source in the GIS FOV has not yet been reported. We tentatively call this source as AX J2049.6+2939. We confirmed it as a compact source within the ASCA imaging capability. The 90 % upper limit of the spatial extent was $14''$. The source location (J2000) is $\alpha = 20^h49^m35^s\!.7$ and $\delta = 29^d38'57''$ with error radius of $70''$(90 % confidence level). It is about $87'$ away from the center.

*2.2. X-ray spectra*

The total exposure time was 11 ks after the standard screening criteria. As shown in figure 1(b), emission from the shell region of the Cygnus Loop extended over AX J2049.6+2939. We estimated the contamination from the shell region using the ROSAT image assuming the spatial distribution of the X-ray emission from the shell region to be similar between the GIS and the PSPC. We retrieved the ROSAT PSPC archival data through the HEASARC/GSFC Online Service. The sequence number of data sets is 500255. Figure 1(c) shows the PSPC image with the GIS contour map. We should note that the source-1 was the only source detected inside the GIS FOV with the PSPC. The shell emission in upper half of the GIS FOV is generally stronger than that in lower half. The shell emission in the region of AX J2049.6+2939 shown in the PSPC image is relatively weak. We selected the background region for the spectral analysis shown with a rectangular region in figure 1(b), considering the intensity of the shell region and the statistics. The source intensity is about 0.11 c s$^{-1}$/GIS and 0.15 c s$^{-1}$/SIS after subtracting the background. We found no intensity variation during our observation period. Figure 2 shows the spectra extracted from a radius of $3'$ centered on AX J2049.6+2939. There is no clear emission line in the spectrum which is quite contrast to those of other regions of the Cygnus Loop (Miyata 1996). We fitted the data using three models: a thermal bremsstrahlung model, the Raymond & Smith model

(Raymond & Smith 1977), and a power law model. The energy bands we adopted were 0.8–9 keV for the GISs and 0.6–9 keV for the SIS. All models gave us acceptable fits folded with the interstellar absorption feature. Results are summarized in Table 1.

*2.3. Short–term variation*

We searched short–term intensity variation using the GIS data. We sacrificed the timing information in the telemetry (timing bit was set to be zero) since the observation was initially intended to obtain the spectrum in detail. When the timing bit is not assigned, the nominal timing resolutions for high and medium bit rate are 62.5 and 500 ms (Ohashi et al. 1996). However, if the observed target is weak enough not to fill the memory capacity which stores event data before sending to telemetry at the constant rate, the photon arrival time can be estimated with the telemetry rate of 3.91 and 31.25 ms for high and medium bit rate as studied by Hirayama et al. (1996). They clearly showed that the photon arrival time could be determined with the telemetry output rate with offset time of 10.6 ms using the data of PSR B0540-69. Since the count rate of our target is much lower than that of PSR B0540-69, we can safely assume the actual timing resolution to be 3.91 and 31.25 ms for high and medium bit rate. We used the data obtained both with high and medium bit rate data for the temporal analysis, resulting the timing resolution of 31.25 ms (32 Hz). After applying the barycentric correction on photon arrival times, we performed the FFT. We found no coherent pulsation. The upper limits (99% confidence level) on the pulsed fraction for the sinusoidal pulse shape are 21 % (0.8-9 keV) and 31 % (2-9 keV), respectively.

*2.4. Other X-ray Observations*

We obtained the Einstein/IPC data set in the vicinity of AX J2049.6+2939 through the HEASARC/GSFC Online Service as the observation ID of 3784. Observation was performed in December 1979. AX J2049.6+2939 was detected and intensity was $0.11 \pm 0.01$ c s$^{-1}$. This value was slightly less than that expected ($0.16 \pm 0.02$ c s$^{-1}$) from the spectral analysis described in section 2.1.

The deep mapping on the Cygnus Loop using the ROSAT HRI is under way (Levenson et al. 1997). The mapping data containing AX J2049.6+2939 is now in the archival data set which can be publicly accessed through the HEASARC/GSFC Online Service as the sequence number of 500462. Observation was performed in November 1995. We noticed three compact sources in the HRI FOV shown in figure 1(c) with the black crosses. One of them coincides with the position of AX J2049.6+2939 determined with ASCA. Therefore, we conclude that this HRI source corresponds to AX J2049.6+2939. Judging from the X-ray spectrum obtained with ASCA, the expected count rate was $(8 \pm 2) \times 10^{-2}$ c s$^{-1}$ whereas the observed value was $(1.03 \pm 0.07) \times 10^{-2}$ c s$^{-1}$. The improved source position is $\alpha = 20^h 49^m 35^s\!.5$ and $\delta = 29^d 38' 47''$ with error radius of $10''$ (90 % confidence level).

We found no short-term variation in the ROSAT HRI data set mainly due to its poor statistics (total number of photon is $\simeq$190).

### 2.5. Counterpart of AX J2049.6+2939

We searched a counterpart in other wavelengths using *Skyview* supported by HEASARC/GSFC. There is no radio source in our error box to a limit of 25 mJy at 4850 MHz (NRAO 4850MHz survey; Condon et al. 1994). There is also no radio sources in any other catalogues available in *W3Browse* supported by HEASARC/GSFC. The closest radio source is $7'\!.7$ away from AX J2049.6+2939 in the north-east direction. Looking at the X-ray image in figure 1(b), there is an extended structure in the north-east direction, which might correspond to another X-ray source associated with the radio source.

We retrieved the digitized sky survey (Lasker et al. 1990) image through the *Skyview*. There are a few stellar objects in our error region as shown in figure 3. We carried out the low-dispersion spectroscopy for the brightest source (V=+12.6; HST Guide Star Catalogue) with the CCD spectrograph mounted on the 1.0-m telescope of Bisei Astronomical Observatory on September 10 1997. The spectral resolution derived from the instrumental profile is $\lambda/\Delta\lambda \simeq 1200$ at 6000–6500 Å. The usable wavelength ranges for the two wavelength setting are 3900–7100 Å and 5500–8500 Å, respectively. The medium dispersion spectroscopic observations were also carried out with the Cassegrain spectrograph (Kosugi et al. 1995) attached to the 188 cm telescope of Okayama Astrophysical Observatory on September 9 and 10 in 1997. The detector used was the SITe 512 × 512 CCD with 20 $\mu$m pixels. The spectral resolving power was $\approx$2000 at 6000 Å. The wavelength regions covered were 4600–5400 Å and 6200–6900 Å. The obtained spectra show the presence of clear absorption lines; Ca H, Ca K, G-band, H$\beta$, Mg b, Na D, and H$\alpha$ with a recession velocity of $\sim 0$ km s$^{-1}$ which indicate that it is a G star.

Taking into account the X-ray intensity and its spectral shape, we conclude that this star has nothing to do with AX J2049.6+2939. We need a deeper observation for much fainter sources within the error region. We assume the upper limit as V$\gtrsim 20$ based on the digitized sky survey image. An X-ray / optical (V-band) flux ratio $f_{X(0.1-2.4\text{keV})}/f_V$ is $\gtrsim 100$ where we assume the power law model for the X-ray flux. Correcting for the interstellar extinction to AX J2049.6+2939 gives $f_{X(0.1-2.4\text{keV})}/f_V \gtrsim 25$.

## 3. Discussion and conclusion

Through the observation project of the whole Cygnus Loop with the ASCA GIS, two compact sources were detected above the 5 $\sigma$ level within the X-ray shell. The brightest compact source is AX J2049.6+2939. The other source was identified to be a K5 star.

We estimated the chance probability of finding an unrelated X-ray source with $\gtrsim 5.3 \times 10^{-13}$ erg s$^{-1}$ in 2-10keV, mainly AGN, inside the Cygnus Loop. Based on the Log$N$–Log$S$ relation studied by Hayashida (1989), the probability is 26 %. This high chance probability cannot exclude the hypothesis of the AGN. Possible long-term variability between ROSAT and ASCA also supports the hypothesis of the AGN. If AX J2049.6+2939 is an AGN, the expected B magnitude is $\sim$16–17 based on the study of the correlation between $f_B$ and $f_X$ (Zamorani et al. 1981). Considering the neutral H column we obtained in the power law model, the B magnitude can be estimated to be $\sim$18–17. Furthermore, it should show emission lines in optical region. The optical spectroscopic study for fainter sources is encouraged.

Normal stars have $f_X/f_V \lesssim 1$ (Stocke et al. 1991), which is much lower than that of AX J2049.6+2939. In the case of X-ray binaries, the color index, $\xi = B_0 + 2.5\log f_{X(2-10\text{keV})}$, is usually introduced to characterize their properties, where $B_0$ is the reddening-corrected B magnitude and $f_{X(2-10\text{keV})}$ is the 2-10keV X-ray flux. Assuming the power law model and B magnitude of 20, $\xi$ is $\simeq 18$. This value is much higher than the average value of high-mass X-ray binaries (7–15; van Paradijs, & McClintock 1995). On the other hand, obtained $\xi$ is lower than the average value of low-mass X-ray binaries ($\simeq 22$). Thus, AX J2049.6+2939 is not likely an X-ray binary.

Flare stars show substantial variability in the X-ray region on the time scale of tens of minutes to hours in their high states (Pallavicini, Tagliaferri, & Stella 1990). On the other hand, no activities have been detected in the case of AX J2049.6+2939. In quiescent states of flare stars, the optical luminosity was well correlated with the X-ray luminosity (Agrawal, Rao, & Sreekantan 1986). Optical luminosities of all of their sample were 3 or 4 orders of magnitude larger than those of X-ray luminosity whereas the optical luminosity of AX J2049.6+2939 was 4 orders of magnitude smaller than that of the X-ray luminosity. Therefore, the hypothesis of flare star can be ruled out.

CVs show a kTe of 10keV or higher (Ishida & Fujimoto 1995) which is much higher than that we obtained. On the contrary, kTe of dwarf novae is generally lower than that of magnetic CVs (Mukai & Shiokawa 1993). Furthermore, they usually show Fe-K emission line with an equivalent width of 500 eV or more. We found no emission line of Fe-K (90 % upper limit is 200 eV). Taking into account these facts, AX J2049.6+2939 is not a CV.

Since the spectra are well fitted by a power law model with a photon index of $-2.1 \pm 0.1$, a rotating neutron star can be a plausible candidate for AX J2049.6+2939 (e.g. Makishima et al. 1996; Becker & Trümper 1997). If AX J2049.6+2939 is a neutron star produced in the SN explosion which left the Cygnus Loop, the transverse velocity is estimated to be $\approx 950$ $D_{770pc} t_{20,000yr}^{-1}$ km s$^{-1}$. In this calculation, we assumed the explosion center is ($\alpha = 20^h 49^m 15^s$, $\delta = 30^d 51' 30''$ (1950); Ku et al. 1984) and the shock wave expanded in spherical symmetry. This assumption is plausible because the X-ray emission at the shell region is originated in the ISM and the proper motion of the ISM is small. Lyne & Lorimer (1994) studied the transverse velocity for 99 pulsars and obtained the average value as $450 \pm 90$ km s$^{-1}$. However, the transverse velocity of the PSR 2224+65 in the Guitar Nebula is estimated to be 986 km s$^{-1}$ (Harrison et al. 1993), which is larger than our case. Therefore, the hypothesis that AX J2049.6+2939 is a neutron star produced in the SN explosion which left the Cygnus Loop is an acceptable idea.

However, the ROSAT observation suggests the possibility of a long-term variability of AX J2049.6+2939. Since the apparent difference of source flux between the ASCA and ROSAT observations might be due to the fact that the source spectrum is not a simple power-law function, we tried another model to characterize the source spectrum. Apparent larger flux in the ASCA band suggests an additional component to the power-law function that dominates at above 2-3 keV. Thus we employed a blackbody component in addition to the power-law function. We found it difficult to independently constrain both the blackbody temperature and the power-law index. Therefore, we assumed the blackbody temperature between 0.1 and 1 keV and estimated the counting rate as observed by the ROSAT HRI. We found that an addition of a blackbody component of temperature $kT \simeq 0.3$keV and emitting radius of $R = 0.1 d_{0.7 \rm kpc}$km slightly decreased the predicted counting rate for ROSAT HRI compared to a simple power-law model. Since the inferred radius is much smaller than that for a neutron star radius and the inferred temperature is higher than canonical cooling model, if such a blackbody component really exists, it is naturally interpreted as a heated polar cap (e.g., Greiveldinger et al. 1996). There was still about a factor of 3 difference between the predicted and observed counting rate for ROSAT HRI and we concluded that the source was variable. This suggests that the source, AX J2049.6+2939 is not an ordinary rotation powered pulsar such as one in the Crab Nebula.

It is well known that an X-ray pulsar 1E2259+586 in the supernova remnant G109.1-1.0 shows a factor of a few flux variation on timescales of a few years (Corbet et al. 1995). Therefore, if the source AX J2049.6+2939 is indeed a young neutron star associated with the Cygnus Loop, we suggest that it might be an anomalous object such as one in CTB 109. We should note, however, that an anomalous pulsar 1E2259+586 has much softer spectral shape compared to that of AX J2049.6+2939.

We reported here the discovery of a compact source within the Cygnus Loop. Based only on the ASCA observational results and optical studies of our error region, we cannot identify AX J2049.6+2939 whether it is an AGN or a neutron star. The obtained X-ray spectrum suggests both possibilities whereas the possible long-term variability supports the hypothesis of an AGN. However, optical studies of our error region prefers the hypothesis of a neutron star. It is strongly required to perform the follow-up observation in various wavelengths in order to reveal its nature. The deeper optical observations for other objects inside the error region will clarify whether they are AGNs or not. In the X-ray regions, we expect the pulsation if it is a rotating neutron star. Based on the work done by Seward & Wang (1988), we can estimate the rate of the rotational energy loss from the X-ray luminosity to be $10^{35}$erg s$^{-1}$. If we assume the characteristic age of AX J2049.6+2939 to be the same as that of the Cygnus Loop, we expect the pulse period to be ~400 ms. Further observation with the ASCA GIS will strongly constrain the pulsed fraction.

We thank Profs. S. Kitamoto and K. Hayashida for valuable comments and suggestions. Dr. B. Aschenbach kindly gave us the whole X-ray image of the Cygnus Loop obtained with the ROSAT all-sky survey. We are grateful to all the members of ASCA team for their contributions to the fabrication of the apparatus, the operation of ASCA, and the data acquisition. Part of this research has made use of data obtained through the High Energy Astrophysics Science Archive Research Center Online Service, provided by the NASA/Goddard Space Flight Center.

Fig. 1. (a) X-ray surface brightness map of the Cygnus Loop obtained with the ROSAT all-sky survey (Aschenbach 1994). The GIS FOV is shown by a black circle.

(b) X-ray image obtained with the ASCA GIS. The GIS image was smoothed with a Gaussian of $\sigma = 1'$. 2–100 % of the peak brightness is logarithmically divided into 15 levels. The rectangular region shows the area accumulated the background in the spectral analysis. (c) X-ray image obtained with the ROSAT PSPC. The contour map is same as (b). Three black crosses correspond to the sources detected with the ROSAT HRI.

Fig. 2. X-ray spectra obtained with GIS-2, GIS-3, and SIS-1. Solid lines show the best fit curves of a power law model. Lower panel shows the residuals of the fits.

Fig. 3. The ROSAT HRI contour map superimposed on the digitized sky survey image. The HRI image was smoothed with a Gaussian of $\sigma = 16''$. AX J2049.6+2939 is at the center of this image and innermost contour level roughly corresponds to the error circle determined with the HRI image.

Table 1. Fitting results

| Model | $\chi^2$/d.o.f | Parameters | $N_{\rm H}$ [$10^{21}{\rm cm}^{-2}$] |
|---|---|---|---|
| Bremsstrahlung | 207.1/194 | kTe = $4.6 \pm 0.5$ | $1.3 \pm 0.5$ |
| Raymond & Smith | 201.7/193 | kTe = $4.3 \pm 0.5$, Z[a] = $0.2 \pm 0.1$ | $1.5 \pm 0.5$ |
| Power law | 191.1/194 | $\Gamma = -2.1 \pm 0.1$ | $3.1 \pm 0.6$ |

NOTED – Quoted errors are at 90% confidence level.

[a] Abundance of heavy elements relative to the cosmic value

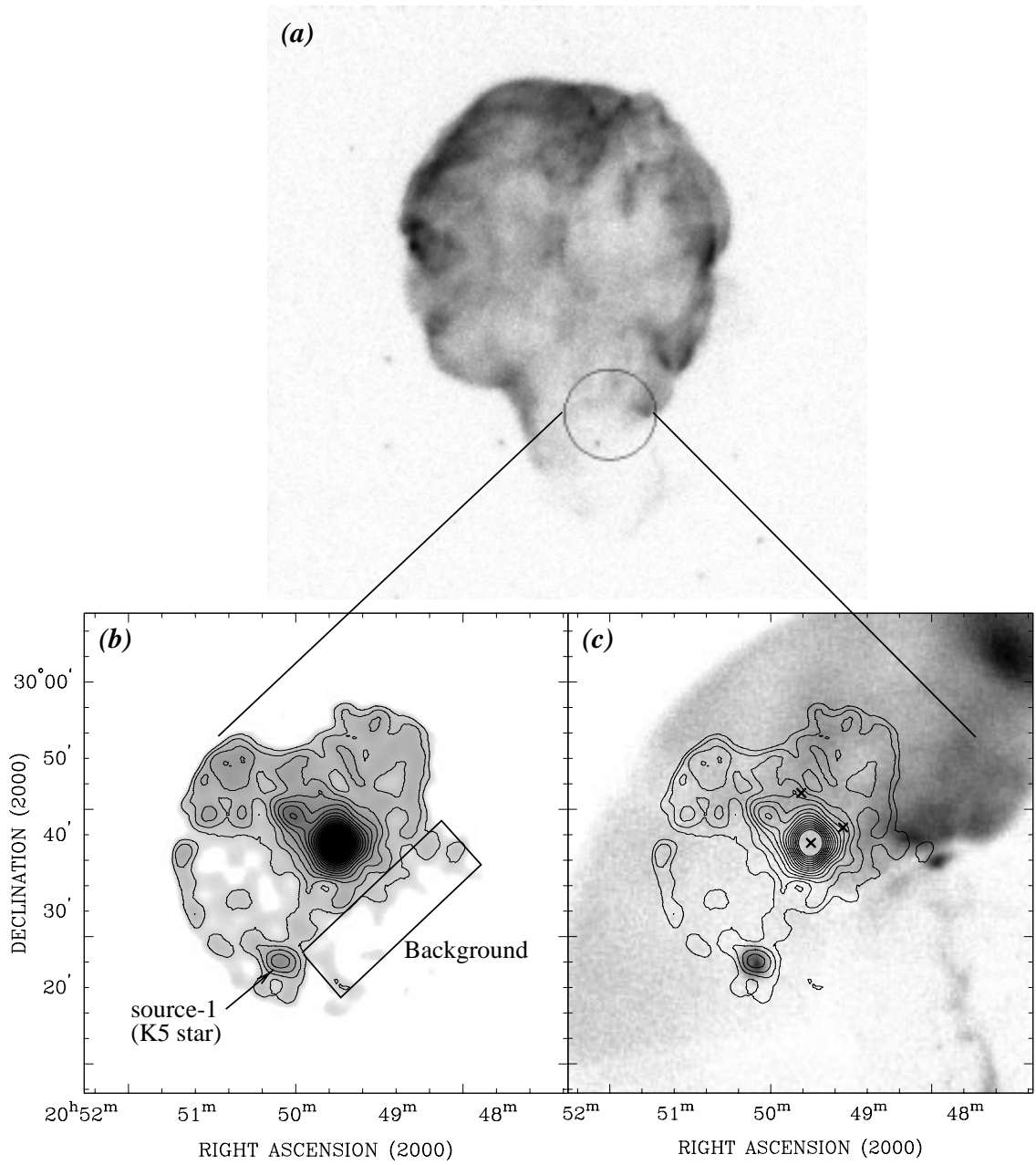

1616

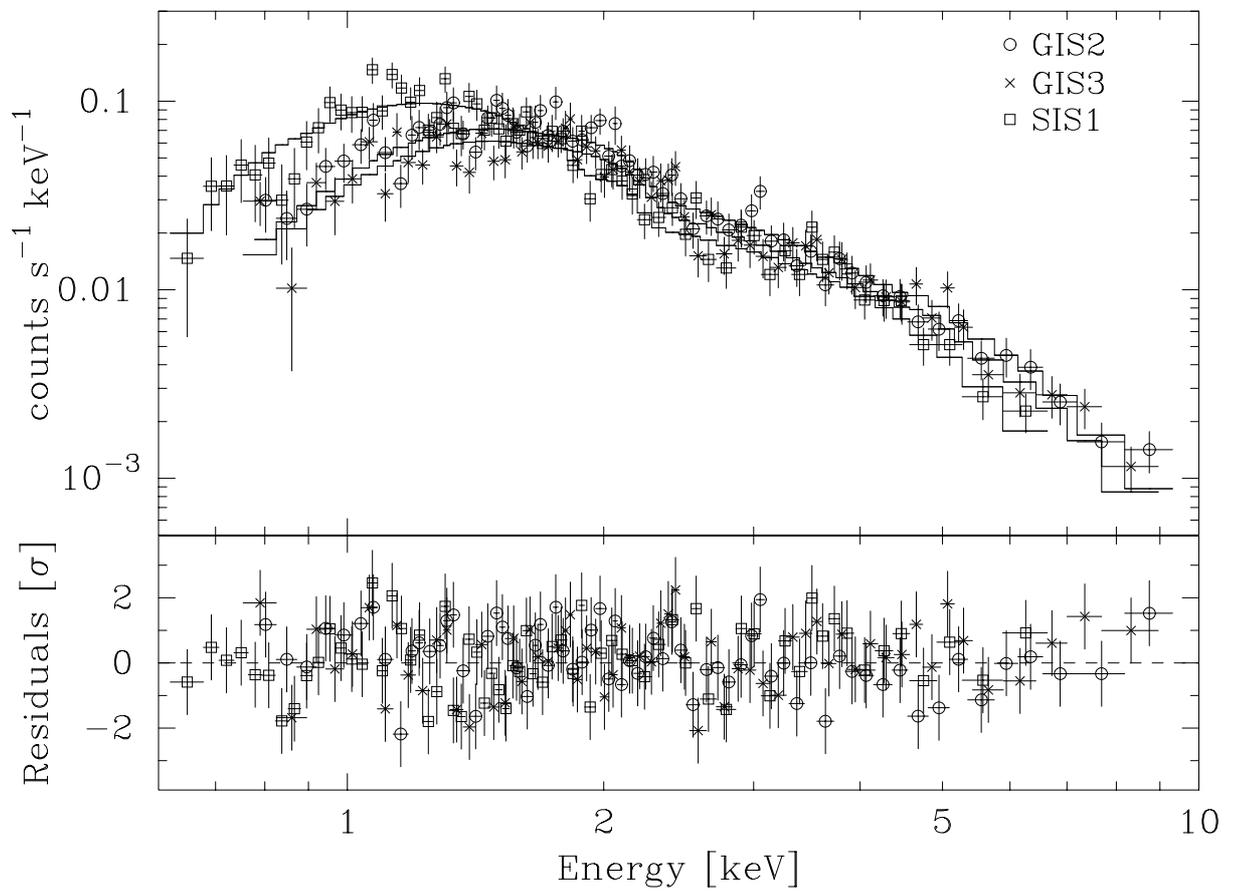

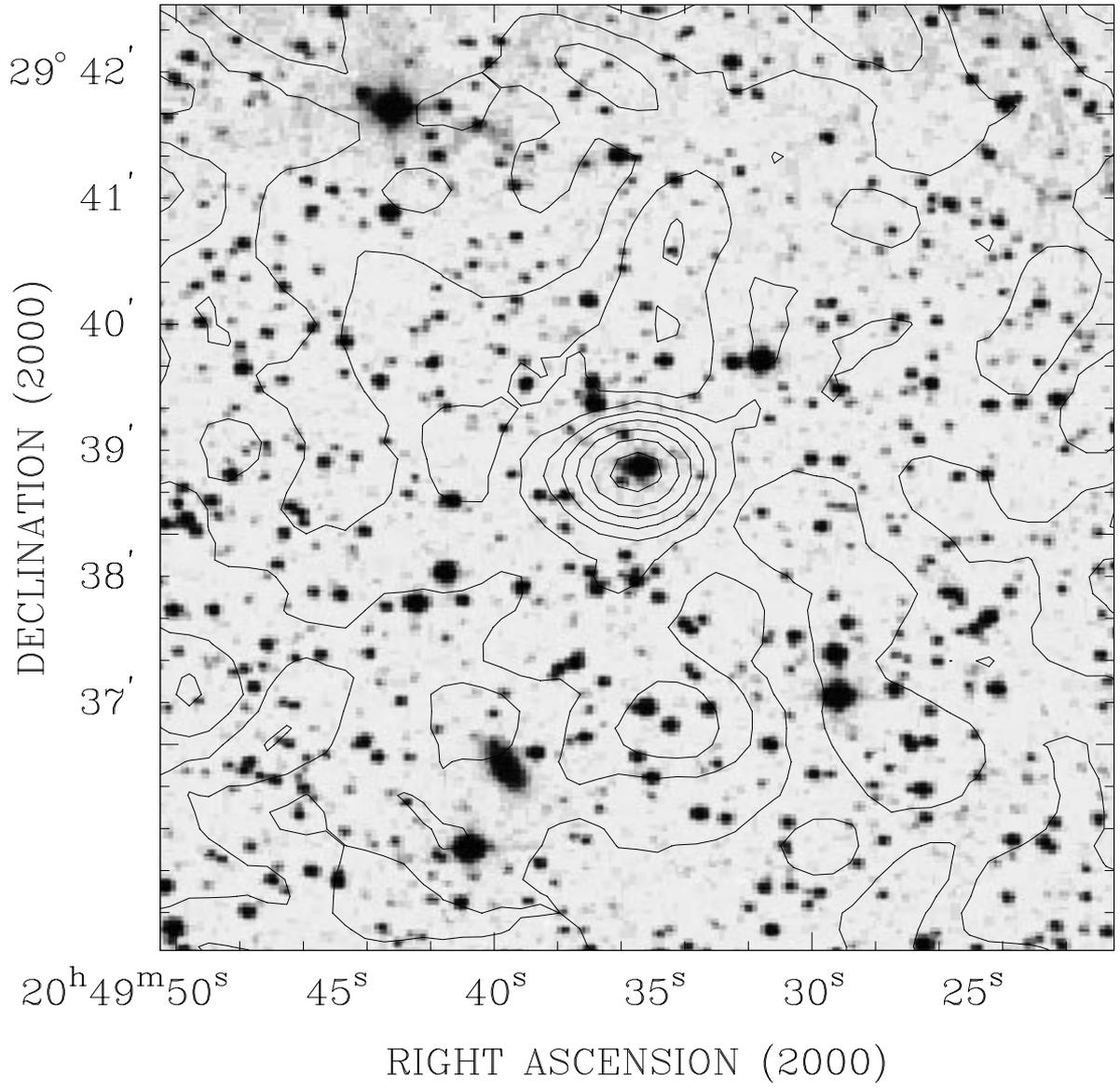